# Observation of the logarithmic dispersion of high frequency edge excitations

N. Q. Balaban, U. Meirav, Hadas Shtrikman and V. Umansky

*Braun Center for Submicron Research, Department of Condensed Matter Physics*
*Weizmann Institute of Science, Rehovot 76100, Israel*

Abstract

We report phase sensitive measurements of microwave propagation through a high mobility two dimensional electron gas subjected to a perpendicular magnetic field. Two types of configurations were used, one that allows all wave vectors $q$, and one that selects only specific $q$ values. The spectrum of edge excitations was studied over a broad frequency span, which allowed us to observe the logarithmic dispersion of edge magneto-plasmons.

Recently, the dynamic excitations of a two dimensional electron gas (2DEG) subjected to a perpendicular magnetic field, *B*, have attracted much attention. Experiments measuring the absorption [1-4] or the transmission [5-7] of electromagnetic waves have detected two main modes evolving from the *B*=0 plasmon. The higher frequency mode corresponds to a bulk excitation of the electrons, whereas the lower frequency mode has been shown to propagate along the edge of the sample [8, 9] and is therefore called the edge magneto-plasmon (EMP).

In a thorough analysis [10], Volkov and Mikhailov (VM) have derived the dispersion relation for the EMP mode,

$$\omega = \frac{\sigma_{xy}(\omega, B)}{2\pi\varepsilon_0\varepsilon^*} q \left[ \ln\left(\frac{2}{ql}\right) + 1 \right] \quad (1)$$

where

$$l = i\frac{\sigma_{xx}(\omega, B)}{2\omega\varepsilon_0\varepsilon^*} \quad (2)$$

is a complex quantity whose magnitude represents the physical extent of the EMP away from the edge. Here $\sigma_{xx}$ and $\sigma_{xy}$ are the components of the conductivity tensor $\sigma$, $\varepsilon^*$ is the effective dielectric constant of the propagation medium and $\varepsilon_0$ the vacuum permittivity. The wave vector of the excitation is *q*, while its angular frequency is Re($\omega$), and Im($\omega$) accounts for the damping. Eq. (1) has been used to describe experimental data for 2DEGs on liquid helium [11] as well as in heterostructures [3]. Most of the experiments were restricted to a fixed *q* which is determined by the perimeter *P* of a mesa, i.e. $q = \frac{2\pi}{P}$. The logarithmic dependence is then reduced to the parameter *l*, which follows the Shubnikov-de-Haas (SdH) oscillations of $\sigma_{xx}$ versus *B* [8, 12]. However, very few experiments have actually traced a logarithmic dependence, as this would require a variation of the parameter $ql$ over a wide range. At relatively low frequency, Grodnenski *et al.* [13] have found evidence for a logarithmic dependence by varying $\sigma_{xx}$, hence $l$, with temperature. At high frequency, Dahl *et al.* [14] have reported a fixed value for $l$ independent of $B$, in contrast with Eq. (2). The logarithmic term was therefore reduced to a constant prefactor rather than following the SdH oscillations [14].

In this work we report microwave transmission measurements in two different geometries, namely with and without ohmic contacts on the boundary, at frequencies from 100 MHz up to 50 GHz. This wide frequency span allows us to observe the logarithmic dependence predicted in Eq. (1). The combination of high frequencies and very high mobility samples has enabled us to test the validity of the VM theory, when the scattering time $\tau$ is long enough to fulfill $\omega\tau \gg 1$. The results agree well with the theory in this limit. We can thus bridge the gap between previous experiments where the oscillatory behavior of $l$ was observed at low frequency [9] and absent at high frequency [14].

The experiments were performed on GaAs/AlGaAs 2DEGs with carrier concentrations $n_s$ from 0.9 to $3 \times 10^{11}$ cm$^{-2}$ and mobilities from 2 to $8 \times 10^6$ cm$^2$/Vs, at 1.5 K. The samples were patterned in rectangular $164 \times 64$ $\mu$m$^2$ mesas. The high frequency input and output are coupled to the sample through a tapered 50$\Omega$ coplanar transmission line, namely a center conductor flanked by two ground electrodes, shown in the inset of Fig. 2. The coupling to the mesa is done in two different ways: capacitive coupling, which we refer to as geometry C, and ohmic contact coupling, or geometry O. In C, the center conductors stop short of the mesa boundary, creating a weak capacitive link to the sample, which generates charge excitations at the mesa edge. The excitation propagates repeatedly around the perimeter $P$, while partially coupling out to the receiving electrode. This circulating motion results in discrete resonance transmission peaks, as will be detailed below. In O, on the other hand, the center conductors form ohmic contacts to the mesa. There, the receiving electrode absorbs all incoming signal, allowing no multiple interference.

In both schemes, the sample constitutes a large impedance mismatch in this transmission line, especially at high magnetic fields where the resistance of the sample increases well above 50 Ω. This mismatch results in strong reflections of the microwave signal. Part of the incoming power is coupled directly to the output lead due to stray capacitances, and we refer to this unintentional contribution to the transmission as 'crosstalk'. The incoming signal also excites EMPs that propagate along the boundary of the sample towards the receiving electrode, acquiring a phase delay and undergoing some attenuation which is generally quite weak at the frequencies used. The total transmission measured is the sum of the crosstalk, which is essentially unaffected by the variation of $B$, and of the EMP signal, whose amplitude and phase are modulated by $B$.

The samples were mounted on specially designed fixtures that couple between the coplanar transmission lines reaching the mesa and two semi-rigid coaxial cables, which were installed in a pumped liquid $^4$He cryostat. Measurements were performed by sending in a high frequency signal on one line and measuring the signal transmitted through the sample to the other, after several stages of wideband amplification. The excitation source was a broad band synthesizer, and the detection was done either by chopping the source and measuring the lock-in voltage on an RF diode, or by measuring the complex transmission parameters with a vector network analyzer. With the latter, the amplitude as well as the *phase* of the signal could be measured, thus allowing us to extract much more information on the dispersion of the EMPs. Detecting the phase was also essential for subtracting the crosstalk from the raw data. A good estimate of the crosstalk can be obtained from measurements at very high $B$, where the high resistance of the samples all but suppresses the transmission through the 2DEG itself. The high field *complex* signal is therefore subtracted from the data.

We first discuss the results obtained in geometry C, where the transmission resonances can be used to study the EMP dispersion at fixed $q$. In Fig. 1(a) the transmission amplitude of sample C is plotted versus $B$ for a frequency $f = 30$ GHz (solid line) and $f = 23$ GHz (dashed line). It consists of a series of peaks, whose characteristic spacing scales inversely with frequency. To underscore the importance of crosstalk deduction, we show in the inset of Fig. 1(a) the same data before subtracting the crosstalk. The shape of the peaks becomes a symmetric Lorenzian, once the crosstalk is subtracted.

These peaks correspond to resonances of the closed mesa and can be understood as follows. The signal traveling along the boundary from the exciting to the receiving electrode will only partially be absorbed in the latter, and then will continue to circulate along the mesa boundary giving rise to multiple interference. Thus, the transmission of the signal in C will show resonances analogous to a Fabry-Perot interferometer. The total phase acquired over the perimeter $P$ has to be an integer multiple of $2\pi$, which corresponds to the condition on the resonant wave vectors $q_n$,

$$q_n = \frac{2\pi}{P} n \tag{3}$$

where *n* is an integer. This picture is substantiated in Fig. 1(b) by plotting the phase of the transmitted signal, which undergoes a shift of π for each consecutive peak, as expected from the Fabry-Perot picture.

However, although the peaks appear equally spaced in *B*, the distance of the first peak from the origin is significantly smaller than the typical spacing. Furthermore, we can follow the evolution of the position of the first peak with increasing *f*, almost down to *B* = 0. Such a behavior cannot be described by a simple hyperbolic dependence of *f* on *B*, as reported in Ref. [14], but it can be understood in terms of the logarithmic dependence predicted in Eq. (1), as we proceed to explain.

For a given index *n*, *q* is fixed according to Eq. (3). Therefore, by following the peaks while varying *f*, the dispersion relation versus *B* at fixed *q* can be measured. In Fig. 2 we plot the positions of the first and second peaks, corresponding to *n* = 1 and *n* = 2. The solid lines are the solution of Eq. (1) and Eq. (2) in the limit of $\omega\tau \gg 1$ and $\omega_c \gg \omega$, with $\omega_c$ the cyclotron frequency. In this limit, using the ac Drude formula, Eq. (2) yields

$$|l| \approx \frac{\sigma_0}{2\omega_c^2 \tau \varepsilon_0 \varepsilon^*} \qquad (4)$$

with $\sigma_0$ the zero field dc conductivity. The dependence of *f* on *B* therefore follows $f \propto B^{-1}\left[\log(B^2) + \text{const.}\right]$, where the $B^2$ comes from the $\omega_c^2$ term in Eq. (4). The only free parameter is $\varepsilon^*$, whose value is expected to be $\varepsilon^* = \frac{\varepsilon_{GaAs}+1}{2} \sim 6.9$. Its value is found to be ~11, most probably enhanced by the presence of the metal surfaces along the sample [13], which distort the electric field lines.

The agreement between data and theory in Fig. 2 is very good. The logarithmic term elucidates the apparent shift of the peaks to lower $B$, which is seen in the data. We point out that the manifestation of the logarithmic dependence was made possible by the combination of high mobility samples and high frequencies, namely $\omega\tau \gg 1$, which enhances the relative importance of the logarithmic term in Eq. (1) over a wide range of $B$. In order to gain further perspective on the dispersion relation, samples of geometry O were studied. Here, the signal propagates along the boundary only once and is absorbed by the ohmic contact of the receiving electrode. Since there is no multiple interference, this geometry does not select specific $q$ values and we do not expect resonant peaks. The phase delay is $\phi = -qL$ and by inverting the real part of Eq. (1), $q(\omega)$ can be expressed as a solution of

$$q = \frac{\text{Re}(\omega) 2\pi\varepsilon_0 \varepsilon^*}{\sigma_{xy}\left(\ln\left(\frac{2}{q|l|}\right)+1\right)} \qquad (5)$$

with $L \equiv P/2$, namely the distance between the two electrodes.

The dependence of the phase on $B$ at various frequencies is plotted in Fig. 3. The oscillatory behavior, due to the SdH oscillations in $\sigma_{xx}$, is seen on top of an overall slope attributed to $\sigma_{xy}$ in Eq. (5). The latter slope of $\phi$ versus $B$ can be approximated from the dispersion relation in the same way as for geometry C, by taking the Drude conductivity in the limit $\omega\tau \gg 1$ with $\omega_c \gg \omega$. Eq. (5) is then solved iteratively for $q$. The results of the calculations for different frequencies are represented by the dashed curves in Fig. 3; again, the only fitting parameter is the effective dielectric constant, which turns out to be $\varepsilon^* \sim 12$, very close to the value deduced for geometry C.

We now turn to discuss the oscillatory part of the transmission. In Fig. 3 one can see the evolution of the SdH oscillations in $\phi$ from 0.8 GHz to 5 GHz. At low frequency, pronounced $1/B$ oscillations appear over the entire range of $B$, while increasing the frequency washes out the high field oscillations. This behavior is opposite to the effect of heating [15], where the lower field oscillations are the first to disappear, as seen in the inset of Fig. 4.

Similarly to the phase, the amplitude of the transmission also shows SdH oscillations, seen in Fig. 4. This oscillatory behavior, which appears on top of a smooth overall variation with $B$, reflects the damping of the EMP, which depends on $l$ and therefore on $\sigma_{xx}$. Indeed, we have observed low field SdH oscillations of the transmission amplitude at frequencies up to 35 GHz. However, at fields higher than ~ 1.2 T, oscillations in the phase and amplitude are observable only up to 5 GHz. For 5 GHz, $\omega\tau$ ~ 2.5, justifying the use of Eq. (4) in order to estimate $l$ around 1.2 T. The extent of the EMP is found from Eq. (4) to be ~ 0.5 µm, comparable to the width of the edge depletion profile. The VM theory assumes a sharp edge, therefore it is not expected to be valid once this limit is reached. For our experimental conditions this limit is attained in the QHE regime and might therefore be linked to the formation of compressible strips at the edge of the 2DEG [16]. Once the extent of the EMP is fully contained in such a strip [17], the SdH oscillations of the bulk $\sigma_{xx}$ should no longer influence the EMP dispersion.

In conclusion, we have investigated the dispersion of edge excitations in a 2DEG by measuring the complex transmission over a wide range of frequencies. We have studied closed boundary systems, where one observes geometric resonances, as well as edge segments delimited by ohmic contacts which do not select particular $q$ values. We have been able to observe the logarithmic dependence of the dispersion predicted by Volkov and Mikhailov. The data measured in both geometries was found to be in good agreement with theory in the limit of $\omega\tau \gg 1$, as long as the lateral extent of the EMP exceeds the depletion profile width.

We wish to acknowledge useful discussions with D. Orgad, and comments from V. I. Talyanskii and G. Ernst. This work was supported by the Israel Science Foundation founded by the Israeli Academy of Science and Humanities.

**Figure captions**

- Fig. 1: (a) Transmission amplitude of a closed mesa (geometry C) at frequencies $f = 23$ GHz (dashed line) and 30 GHz (solid line), for a sample with mobility $2\times10^6$ cm$^2$/Vs and density $n_s = 2.3\times10^{11}$ cm$^{-2}$. The peaks correspond to different $n$ in Eq. (3). The curves are vertically offset for clarity. (b) The phase of the transmitted signal at $f = 30$ GHz. Each resonance peak corresponds to a decrease of $\pi$, as explained in the text. Inset: The measured data at $f = 30$ GHz before subtraction of the crosstalk, showing the interference of the latter with the EMP signal.

- Fig. 2: Position of peaks with $n=1$ (●) and 2 (◆) in Eq. (3), namely $q=2\pi/P$ and $4\pi/P$, respectively, for a sample with a mobility of $5\times10^6$ cm$^2$/Vs and $n_s=1\times10^{11}$ cm$^{-2}$. The solid lines are solutions of Eq. (1) with $\varepsilon^* = 11$. We remark that the data cannot be fitted without taking into account the logarithmic dependence. Inset: Schematic drawing of the samples used, where the mesa is marked in gray and the metallic surfaces in black. The latter form a tapered coplanar transmission line reaching the sample, where the inner conductor width narrows down to 6 μm. In geometry C, the center conductors couple capacitively to the mesa, whereas in geometry O, they form ohmic contacts. Geometry O has an additional gate patterned on the inner part of the mesa, which helps reduce the crosstalk. Its effect on the EMP dispersion is relevant only for small $q$ or large $l$, and is negligible once the magnetic field exceeds a few tenths of a Tesla.

- Fig. 3: Phase of the transmitted signal in geometry O, for several frequencies. The SdH oscillations persist at high frequencies only in the low *B* regime. The sample parameters are the same as in Fig. 1. The dashed lines result from the iterative solution of Eq. (5), using the Drude form of Eq. (4) for *l*.

-Fig. 4: Transmission amplitude at 0.2 GHz (dashed line) and 18 GHz (solid line), showing SdH oscillations in the attenuation of EMP. The sample, of the same material as in Fig. 1, is patterned in geometry O. Inset: The phase of the transmitted signal at 5 GHz for two different power levels, for the same sample. The curves are vertically offset for clarity. The upper curve corresponds to a power level 10 times higher than the lower curve. Note how heating primarily tends to affect the lower field SdH oscillations.

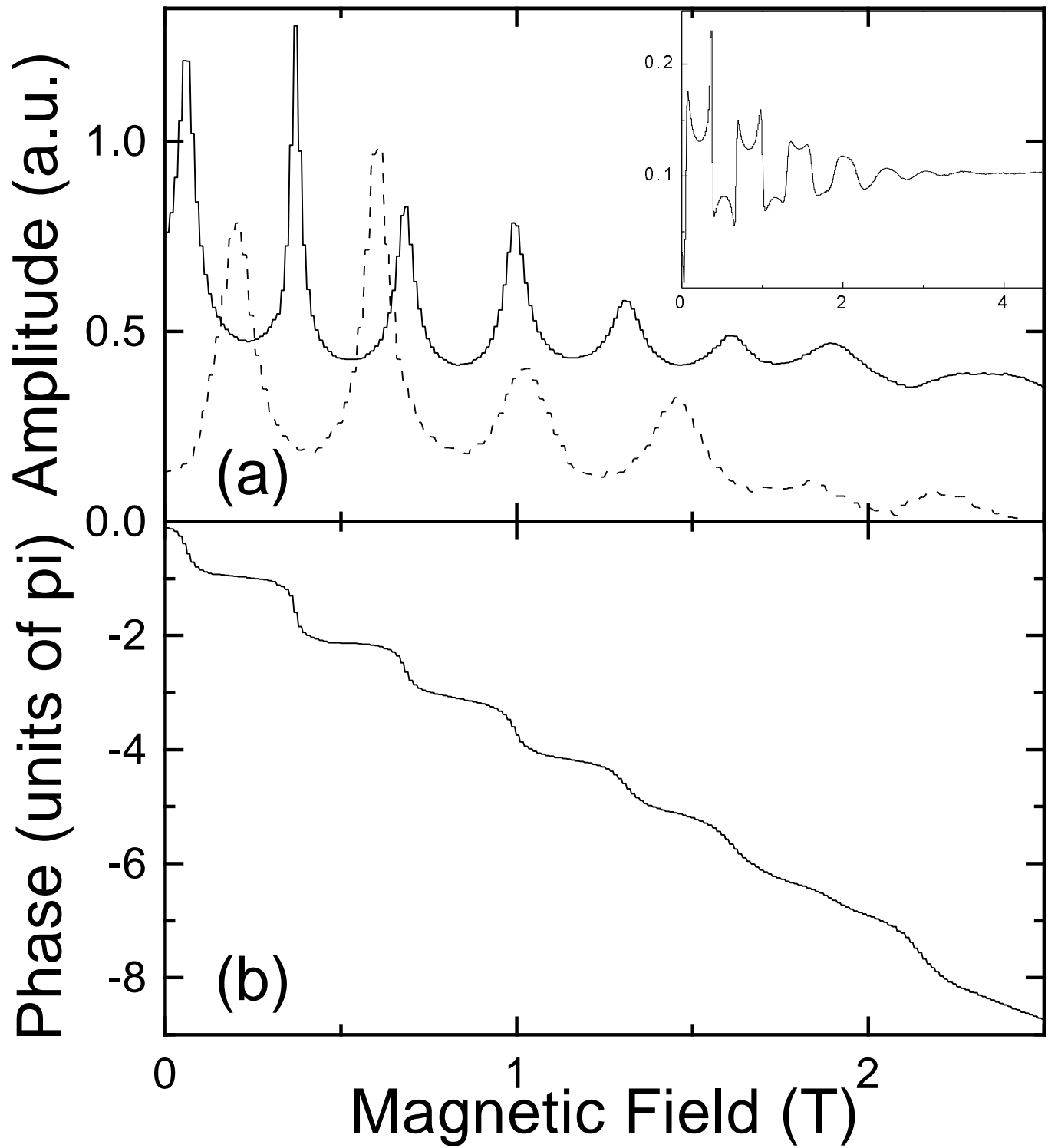

Fig. 1

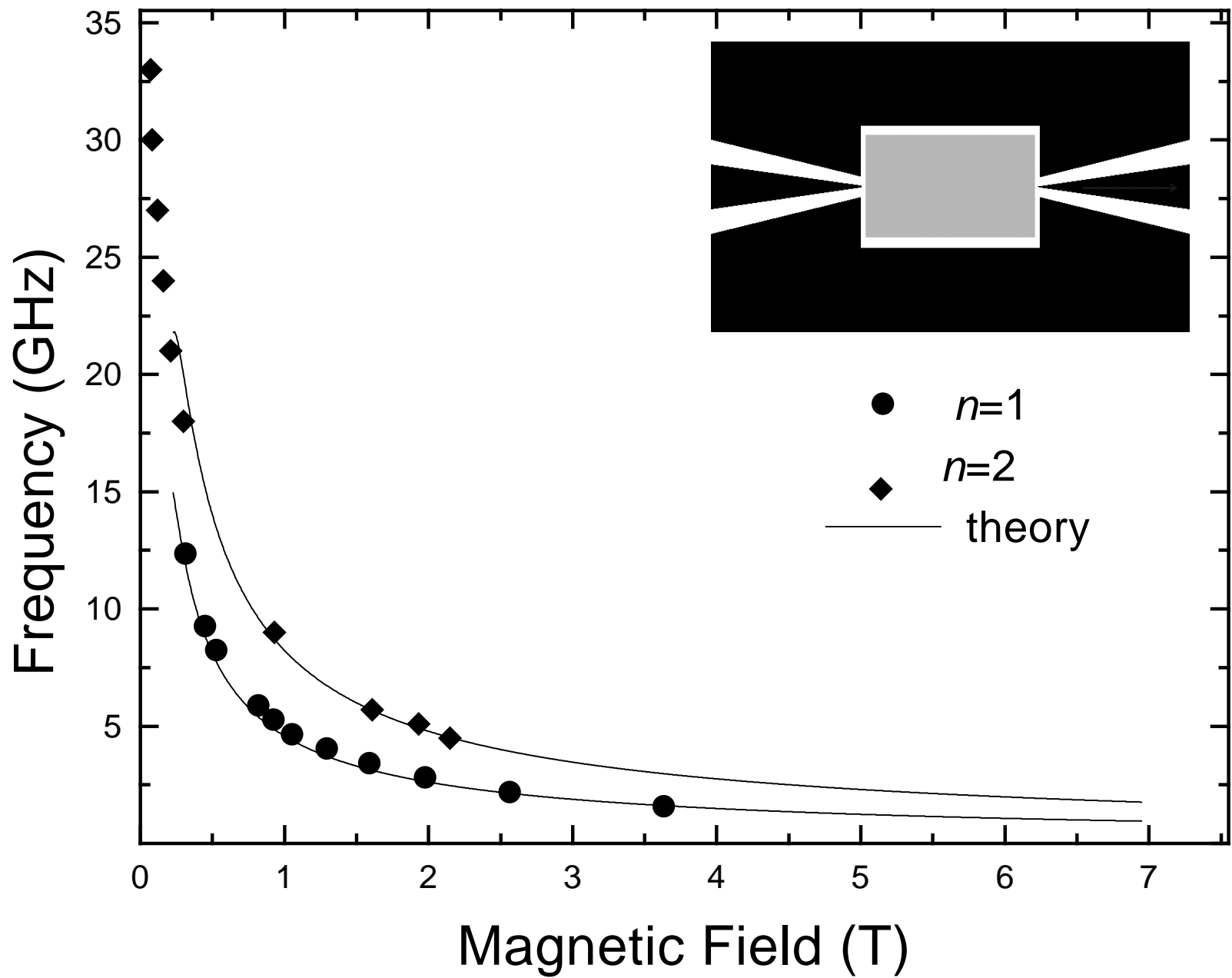

Fig. 2

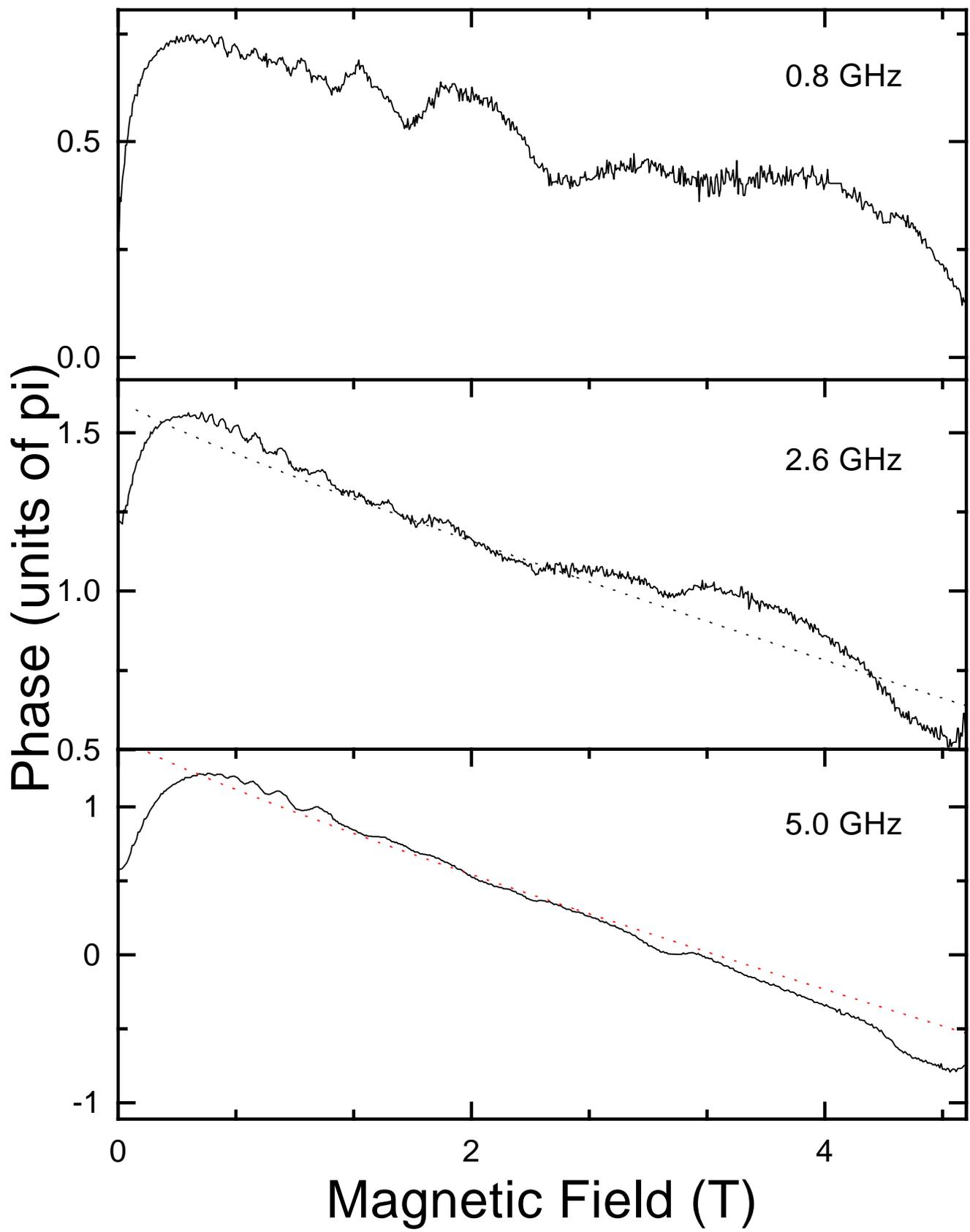

Fig. 3

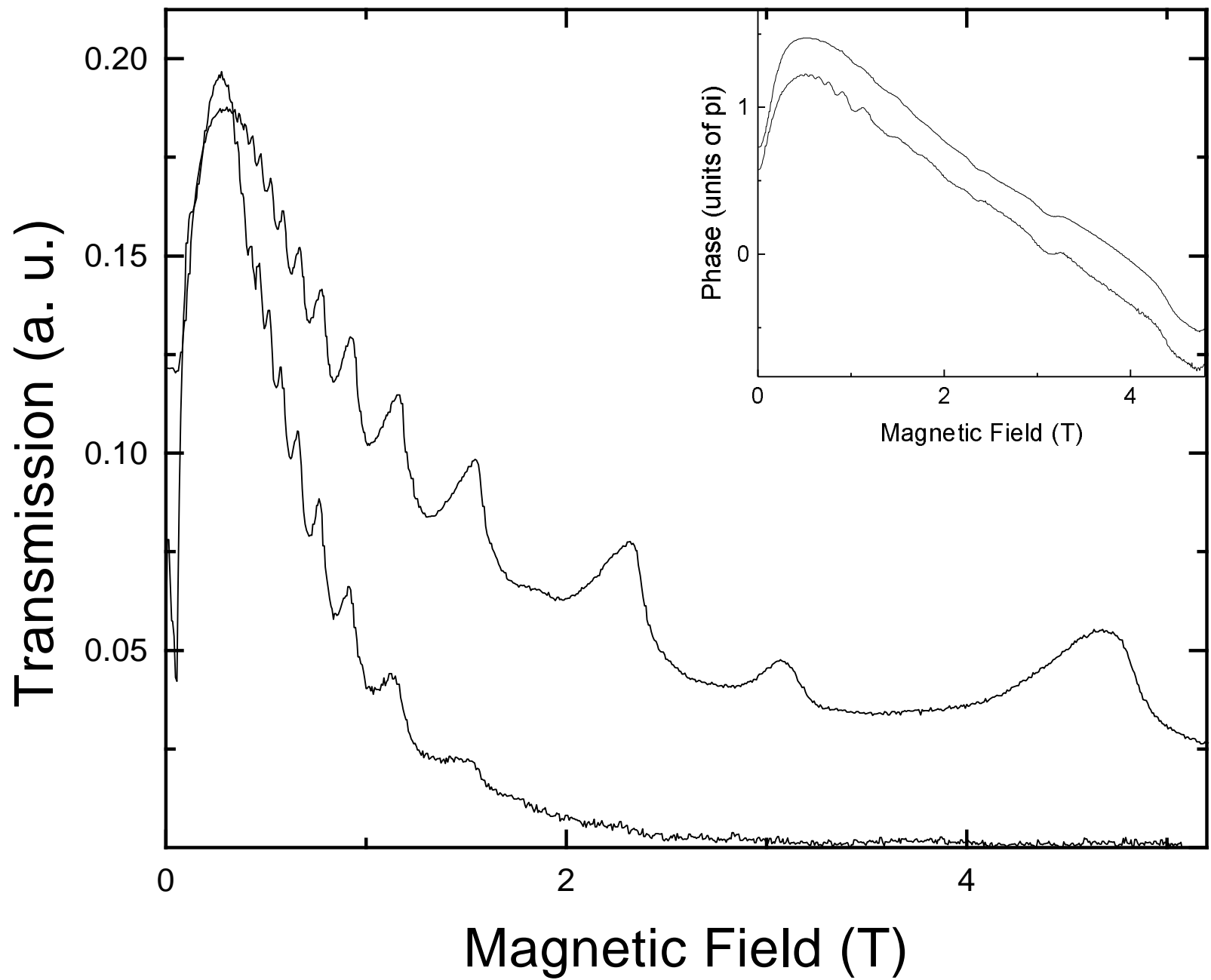

Fig. 4